\documentclass[preprint,10pt,a4paper,twocolumn,final,tightenlines,showpacs,bibnotes,nofootinbib,notitlepage,pra]{revtex4}
\usepackage{amssymb}

\usepackage{amsmath}

\newtheorem{theorem}{Theorem}
\newtheorem{acknowledgement}[theorem]{Acknowledgement}

\input{tcilatex}

\begin{document}

\title{Non-Markov dynamics and phonon decoherence of a double quantum dot
charge qubit}
\author{Xian-Ting Liang\thanks{%
Electronic address: xtliang@ustc.edu}}
\affiliation{Department of Physics and Institute of Modern Physics, Ningbo University,
Ningbo, 315211, China}

\begin{abstract}
In this paper we investigate decoherence times of a double quantum dot (DQD)
charge qubit due to it coupling with acoustic phonon baths. We individually
consider the acoustic piezoelectric as well as deformation coupling phonon
baths in the qubit environment. The decoherence times are calculated with
two kinds of methods. One of them is based on the qusiadiabatic propagator
path integral (QUAPI) and the other is based on Bloch equations, and two
kinds of results are compared. It is shown that the theoretical decoherence
times of the DQD charge qubit are shorter than the experimental reported
results. It implies that the phonon couplings to the qubit play a
subordinate role, resulting in the decoherence of the qubit.
\end{abstract}

\maketitle

\section{Introduction}

Solid state qubits are considered to be promising candidates for realizing
building blocks of quantum information processors because they can be scaled
up to large numbers. The double quantum dot (DQD) \cite%
{DQD1,DQD2,DQD3,DQD4,Brandes} charge qubit is one of these qubits. Two
low-energy charge states are used as the local states $\left| 0\right\rangle 
$ and $\left| 1\right\rangle $ in the qubit. The qubit can be controlled
directly via external voltage sources. There are some effective schemes to
prepare the initial states and read out the final states of the qubit \cite%
{prepare-and-measure}. So it is considered that decoherence may be the
central impediment for the qubit to be taken as the cell of quantum
computer. Finding out the primary origin or the dominating mechanism of
decoherence for the qubit is a basal task for overcoming the impediment.
Experimentally, many attempts \cite{experiment1,experiment2} for detecting
the decoherence time of this kind of qubit have been performed. The
decoherence has also been investigated theoretically. In 2000, Fedichkin 
\emph{et al. }\cite{Fedichkin-etal01} investigated the Born-Markov type
electron-phonon decoherence at large times due to spontaneous phonon
emission of the quantum dot charge qubits. Recently, Vorojtsov \emph{et al.} %
\cite{Vorojtsov-etal} studied the decoherence of the DQD charge qubit by
using Born-Markov approximation. But, as it has been pointed out, the use of
the Born-Markov approximation is inappropriate at large tunneling
amplitudes. The method is expected to become increasingly unreliable at DQD
with larger interdot tunneling amplitudes. Wu \emph{et al.} \cite{Wu-etal}
investigated the decoherence in terms of a perturbation treatment based on a
unitary transformation. The Born-Markov approximation has not been used in
the method but it neglects some terms of the effective Hamiltonian with high
excited states. This kind of processing introduces an new approximation
which has not been estimated to the affects of the dynamics. Fedichkin \emph{%
et al.} \cite{Fedichkin01,Fedichkin02} studied the error rate of DQD charge
qubit with short-time approximation. This method is accurate enough in
adequate short time. But the decoherence in a moderately long time is also
interesting. Recently, Thorwart \emph{et al. }\cite{Thorwartetal2}
investigated the decoherence of the DQD charge qubit in a longer time with a
numerically exact iterative quasiadiabatic propagator path integral (QUAPI) %
\cite{Makri}. This method is proved valid in investigating the qubit
decoherence \cite{Thorwartetal1}. In Ref. \cite{Thorwartetal2}, Thorwart 
\emph{et al.} considered the coupling of longitudinal piezoelectic acoustic
phonons with the investigated qubit and neglected the contribution of the
deformation acoustic phonons to decoherence. These two kinds of phonons may
constitute two kinds of different coupling baths in the environment of the
qubit. We call the former the piezoelectric coupling phonon bath (PCPB) and
the latter the deformation coupling phonon bath (DCPB). Comparing Thorwart's
result\ and the reported experimental value they found that the theory
predicts the decoherence time of the DQD charge qubit is two orders of
magnitudes smaller than the experimental one. Thus, Thorwart \emph{et al. }%
conclude that\emph{\ }the piezoelectric coupling phonon decoherence is a
subordinate mechanism in decoherence of the DQD charge qubit. Recently, Wu 
\emph{et al.} \cite{Wu-etal} gave the spectral density functions of PCPB as
well as DCPB. Then how about the DCPB to the decoherence of the DQD charge
qubit? In other words, is the decoherence of the DQD charge qubit induced by
DCPB also subordinate? In this paper we shall use an iterative tensor
multiplication (ITM) \cite{Makri} scheme derived from the QUAPI to study the
decoherence times of the DQD charge qubit not only in PCPB but also in DCPB.
In order to validate if our result is in accordance with Thorwart's result
we at first investigate the decoherence times of the qubit in PCPB. Then we
shall investigate the decoherence times of the qubit in another bath, DCPB,
which will show that the influence of the DCPB to the decoherence of the DQD
charge qubit is also subordinate because it results in a shorter decoherence
time than the experimental value of $1$ $ns$ \cite{experiment1,experiment2}.

\section{Models}

The DQD charge qubit consists of left and right dots connected through an
interdot tunneling barrier. Due to Coulomb blockade, at most one excess
electron is allowed to occupy the left and right dot, which defines two
basis vectors $\left| 0\right\rangle $ and $\left| 1\right\rangle .$ The
energy difference $\varepsilon $ between these two states can be controlled
by the source-drain voltage. Neglecting the higher order tunneling between
leads and the dots, the effective Hamiltonian in the manipulation process
reads \cite{Wu-etal,Thorwartetal2}%
\begin{equation}
H_{eff}=\hbar T_{c}\sigma _{x}+\hbar \sum_{q}\omega _{q}b_{q}^{\dagger
}b_{q}+\hbar \sigma _{z}\sum_{q}\left( M_{q}b_{q}^{\dagger }+M_{q}^{\ast
}b_{q}\right) .  \label{eq1}
\end{equation}%
Here, $T_{c}$ is the interdot tunneling, $\sigma _{x}$ and $\sigma _{z}$ are
Pauli matrix, $b_{q}^{\dagger }$ $(b_{q})$ are the creation (annihilation)
operators of phonons, $\hbar \omega _{q}$ is the energy of the phonons, and $%
M_{q}=C_{q}/\sqrt{2m_{q}\omega _{q}\hbar }$ where $C_{q}$ are the classical
coupling constants of qubit-phonon system. We call the collective coupling
phonons to the qubit in the environment a phonon bath. In order to obtain
the reduced density matrix of the qubit in the system, one should know the
coupling coefficients $M_{q}$, but in fact we need not know the details of
each $M_{q}$ because all characteristics of the bath pertaining to the
dynamics of the observable system are captured in the spectral density
function \cite{Weiss,Leggett}%
\begin{equation}
J\left( \omega \right) =\sum_{q}\left| M_{q}\right| ^{2}\delta \left( \omega
-\omega _{q}\right) .  \label{eq2}
\end{equation}%
It is pointed out that the spectral density of PCPB is%
\begin{equation}
J^{pz}\left( \omega \right) =g_{pz}\omega \left( 1-\frac{\omega _{d}}{\omega 
}\sin \frac{\omega }{\omega _{d}}\right) e^{-\frac{\omega ^{2}}{2\omega
_{l}^{2}}}.  \label{eq3}
\end{equation}%
Here, $\omega _{d}=s/d$ and $\omega _{l}=s/l$, where $d$ denotes the
center-to-center distance of two dots, $l$ the dot size, $s$ the sound
velocity in the crystal, and 
\begin{equation*}
g_{pz}=\frac{M}{\pi ^{2}\varrho s^{3}}\left( \frac{6}{35}+\frac{1}{x}\frac{8%
}{35}\right) .
\end{equation*}%
Here, $M$ is the piezoconstant, $\varrho $ is the density of the crystal,
and $x$ is the rate of transverse to the longitudinal of sound velocity in
the crystal, (see for example Refs. \cite{Wu-etal,Fedichkin01}). As in Refs. %
\cite{Wu-etal,Fedichkin01} in this paper we set the sound velocity in the
GaAs crystal $s\approx 5\times 10^{3}$ $m/s$. With the parameters of GaAs in
Ref. \cite{Mahan-etal}, Wu \emph{et al.} \cite{Wu-etal} propose a value $%
g_{pz}\approx 0.035$ $(ps)^{-2}$. The spectral density of DCPB is 
\begin{equation}
J^{df}\left( \omega \right) =g_{df}\omega ^{3}\left( 1-\frac{\omega _{d}}{%
\omega }\sin \frac{\omega }{\omega _{d}}\right) e^{-\frac{\omega ^{2}}{%
2\omega _{l}^{2}}},  \label{eq4}
\end{equation}%
where 
\begin{equation*}
g_{df}=\frac{\Xi ^{2}}{8\pi ^{2}\varrho s^{5}}.
\end{equation*}%
Here, $\Xi $ is the deformation potential. In the same paper, Wu \emph{et al.%
} also propose a value $g_{df}\approx 0.029$ $(ps)^{-2}.$ One can
investigate the dynamics and then the decoherence of the open qubit with the
help of the definite spectral density functions of the baths. Before
investigations of decoherence of the DQD charge qubit we introduce an
optimal numerical path integral method, the ITM method in the following
section.

\section{QUAPI and ITM}

In the following, we firstly review the QUAPI and then the ITM \cite{Makri}
scheme. Suppose the initial state of the qubit-bath system has the form%
\begin{equation}
R\left( 0\right) =\rho \left( 0\right) \otimes \rho _{bath}\left( 0\right) ,
\label{eq6}
\end{equation}%
where $\rho \left( 0\right) $ and $\rho _{bath}\left( 0\right) $ are the
initial states of the qubit and bath. The evolution of the reduced density
operator of the open qubit%
\begin{equation}
\tilde{\rho}\left( s^{\prime \prime },s^{\prime };t\right) =\text{Tr}%
_{bath}\left\langle s^{\prime \prime }\right| e^{-iHt/\hbar }\rho \left(
0\right) \otimes \rho _{bath}\left( 0\right) e^{iHt/\hbar }\left| s^{\prime
}\right\rangle ,  \label{eq7}
\end{equation}%
is given by%
\begin{eqnarray}
&&\tilde{\rho}\left( s^{\prime \prime },s^{\prime };t\right)   \notag \\
&=&\sum_{s_{0}^{+}=\pm 1}\sum_{s_{1}^{+}=\pm 1}\cdot \cdot \cdot
\sum_{s_{N-1}^{+}=\pm 1}\sum_{s_{0}^{-}=\pm 1}\sum_{s_{1}^{-}=\pm 1}\cdot
\cdot \cdot \sum_{s_{N-1}^{-}=\pm 1}  \notag \\
&&\times \left\langle s^{\prime \prime }\right| e^{-iH_{0}\Delta t/\hbar
}\left| s_{N-1}^{+}\right\rangle \cdot \cdot \cdot \left\langle
s_{1}^{+}\right| e^{-iH_{0}\Delta t/\hbar }\left| s_{0}^{+}\right\rangle  
\notag \\
&&\times \left\langle s_{0}^{+}\right| \rho \left( 0\right) \left|
s_{0}^{-}\right\rangle   \notag \\
&&\times \left\langle s_{0}^{-}\right| e^{iH_{0}\Delta t/\hbar }\left|
s_{1}^{-}\right\rangle \cdot \cdot \cdot \left\langle s_{N-1}^{-}\right|
e^{iH_{0}\Delta t/\hbar }\left| s^{\prime }\right\rangle   \notag \\
&&\times I\left( s_{0}^{+},s_{1}^{+},\cdot \cdot \cdot
,s_{N-1}^{+},s^{\prime \prime },s_{0}^{-},s_{1}^{-},\cdot \cdot \cdot
,s_{N-1}^{-},s^{\prime };\Delta t\right) ,  \notag \\
&&  \label{eq8}
\end{eqnarray}%
where the influence functional is%
\begin{eqnarray}
&&I\left( s_{0}^{+},s_{1}^{+},\cdot \cdot \cdot ,s_{N-1}^{+},s^{\prime
\prime },s_{0}^{-},s_{1}^{-},\cdot \cdot \cdot ,s_{N-1}^{-},s^{\prime
};\Delta t\right)   \notag \\
&=&\text{Tr}_{bath}\left[ e^{-iH_{env}\left( s^{\prime \prime }\right)
\Delta t/2\hbar }e^{-iH_{env}\left( s_{N-1}^{+}\right) \Delta t/2\hbar
}\right.   \notag \\
&&\times \cdot \cdot \cdot e^{-iH_{env}\left( s_{0}^{+}\right) \Delta
t/2\hbar }\rho _{bath}\left( 0\right) e^{iH_{env}\left( s_{0}^{-}\right)
\Delta t/2\hbar }  \notag \\
&&\left. \times \cdot \cdot \cdot e^{iH_{env}\left( s_{N-1}^{-}\right)
\Delta t/2\hbar }e^{iH_{env}\left( s\prime \right) \Delta t/2\hbar }\right] .
\label{eq9}
\end{eqnarray}%
Here, $H_{0}$ is a reference Hamiltonian that in general depends on the
coordinate and momentum of the system. In the qubit system, it usually
depends on Pauli matrixes $\sigma _{x}$ and $\sigma _{z}$. The $H_{env}$ is
defined as $H_{env}=H-H_{0}$. In our system we set $H_{0}=\hbar T_{c}\sigma
_{x}.$ The discrete path integral representation of the qubit density matrix
contains temporal nonlocal terms $I\left( s_{0}^{+},s_{1}^{+},\cdot \cdot
\cdot ,s_{N-1}^{+},s^{\prime \prime },s_{0}^{-},s_{1}^{-},\cdot \cdot \cdot
,s_{N-1}^{-},s^{\prime };\Delta t\right) $ which denotes the process being
non-Markovian. With the quasiadiabatic discretization of the path integral,
the influence functional, Eq.(\ref{eq9}) takes the form%
\begin{equation}
I=\exp \left\{ -\frac{i}{\hbar }\tsum_{k=0}^{N}\tsum_{k^{\prime
}=0}^{k}\left( s_{k}^{+}-s_{k}^{-}\right) \left( \eta _{kk^{\prime
}}s_{k^{\prime }}^{+}-\eta _{kk^{\prime }}^{\ast }s_{k^{\prime }}^{-}\right)
\right\} ,  \label{eq10}
\end{equation}%
where $s_{N}^{+}=s^{\prime \prime }$ and $s_{N}^{-}=s^{\prime }.$ The
coefficients $\eta _{kk^{\prime }}$ can be obtained by substituting the
discrete path into the Feynman-Vernon expression. Their expressions have
been shown in Ref. \cite{Makri}. Thus, the influence functional can be
expressed with a product of terms corresponding to different $\Delta k$ as 
\begin{eqnarray}
I &=&\tprod_{k=0}^{N}I_{0}\left( s_{k}^{\pm }\right)
\tprod_{k=0}^{N-1}I_{1}\left( s_{k}^{\pm },s_{k+1}^{\pm }\right)
\tprod_{k=0}^{N-\Delta k}I_{\Delta k}\left( s_{k}^{\pm },s_{k+\Delta k}^{\pm
}\right)   \notag \\
&&...\tprod_{k=0}^{N-\Delta k_{\max }}I_{\Delta k_{\max }}\left( s_{k}^{\pm
},s_{k+\Delta k_{\max }}^{\pm }\right) .  \label{eq11}
\end{eqnarray}%
Here, $\Delta k=k-k^{\prime },$ where $k^{\prime }$ and $k$ are points of
discrete path integral expressions, (see Ref. \cite{Makri}) and%
\begin{eqnarray}
I_{0}\left( s_{i}^{\pm }\right)  &=&\exp \left\{ -\frac{1}{\hbar }\left(
s_{i}^{+}-s_{i}^{-}\right) \left( \eta _{ii}s_{i}^{+}-\eta _{ii}^{\ast
}s_{i}^{-}\right) \right\} ,  \notag \\
I_{\Delta k}\left( s_{i}^{\pm },s_{i+\Delta k}^{\pm }\right)  &=&\exp
\left\{ -\frac{1}{\hbar }\left( s_{i+\Delta k}^{+}-s_{i+\Delta k}^{-}\right)
\right.   \notag \\
&&\times \left. \left( \eta _{i+\Delta k,i}s_{i}^{+}-\eta _{i+\Delta
k,i}^{\ast }s_{i}^{-}\right) \right\} ,\Delta k\geqslant 1.  \notag \\
&&  \label{eq12}
\end{eqnarray}%
The length of the memory of the time can be estimated by the following bath
response function%
\begin{equation}
\alpha ^{x}\left( t\right) =\frac{1}{\pi }\int_{0}^{\infty }d\omega
J^{x}\left( \omega \right) \left[ \coth \left( \frac{\beta \hbar \omega }{2}%
\right) \cos \omega t-i\sin \omega t\right] .  \label{eq13}
\end{equation}%
Here, the superscript $x$ denotes the bath type, $\beta =1/k_{B}T,$ where $%
k_{B}$ is the Boltzmann constant, and $T$ is the temperature. It is shown
that when the real and imaginary parts behave as the delta function $\delta
\left( t\right) $ and its derivative $\delta ^{\prime }\left( t\right) ,$
the dynamics of the reduced density matrix is Markovian. However, if the
real and imaginary parts are broader than the delta function, the dynamics
is non-Markovian. The broader the Re$[\alpha ^{x}\left( t\right) ]$ and Im$%
[\alpha ^{x}\left( t\right) ]$ are, the longer of the memory time will be.
The broader the Re$[\alpha ^{x}\left( t\right) ]$ and Im$[\alpha ^{x}\left(
t\right) ]$ are, the more serious the Markov approximation will distort the
practical dynamics. In Fig.1 we plot the Re$[\alpha ^{pz}\left( t\right) ]$
and Im$[\alpha ^{pz}\left( t\right) ]$ of the PCPB and in Fig.2 we plot the
Re$[\alpha ^{df}\left( t\right) ]$ and Im$[\alpha ^{df}\left( t\right) ]$ of
the DCPB.%
\begin{eqnarray*}
&& \\
&& \\
&&Fig.1, \\
&& \\
&& \\
&&Fig.2 \\
&& \\
&&
\end{eqnarray*}%
We see that the memory times are about $\tau _{mem}^{pz}=1\times 10^{-11}$ $s
$ for PCPB and $\tau _{mem}^{df}=2\times 10^{-11}$ $s$ for DCPB \{where the
points beyond $\pm 1\times 10^{-11}$ $s$ have not been plotted for clearly
distinguishing the Re$[\alpha ^{df}\left( t\right) ]$ and Im$[\alpha
^{df}\left( t\right) ]$ in the same figure\}. Due to the nonlocality, it is
impossible to calculate the reduced density matrix by Eq.(\ref{eq8}) in the
matrix multiplication scheme. However, the short range nonlocality of the
influence functional implies that the effects of the nonlocality should drop
off rapidly as the ``interaction distance''\ increases. In the ITM scheme
the interaction can be taken into account at each iteration step. The
reduced density matrix at time $t=N\Delta t$ ($N$ even) is given as%
\begin{equation*}
\tilde{\rho}\left( s_{N}^{\pm },N\Delta t\right) =A^{\left( 1\right) }\left(
s_{N}^{\pm };N\Delta t\right) I_{0}\left( s_{N}^{\pm }\right) ,
\end{equation*}%
where%
\begin{eqnarray*}
A^{\left( 1\right) }\left( s_{k+1}^{\pm };(k+1)\Delta t\right)  &=&\int
ds_{k}^{\pm }T^{\left( 2\right) }\left( s_{k}^{\pm },s_{k+1}^{\pm }\right) 
\\
&&\times A^{\left( 1\right) }\left( s_{k}^{\pm };k\Delta t\right) .
\end{eqnarray*}%
Here,%
\begin{eqnarray*}
&&T^{\left( 2\Delta k_{\max }\right) }\left( s_{k}^{\pm },s_{k+1}^{\pm
}...s_{k+2\Delta k_{\max }-1}^{\pm }\right)  \\
&=&\tprod_{n=k}^{k+\Delta k_{\max }-1}K\left( s_{k}^{\pm },s_{k+1}^{\pm
}\right) I_{0}\left( s_{n}^{\pm }\right) I_{1}\left( s_{n}^{\pm
},s_{n+1}^{\pm }\right)  \\
&&\times I_{2}\left( s_{n}^{\pm },s_{n+2}^{\pm }\right) ...I_{\Delta k_{\max
}}\left( s_{n}^{\pm },s_{n+\Delta k_{\max }}^{\pm }\right) ,
\end{eqnarray*}%
and%
\begin{equation*}
A^{\left( \Delta k_{\max }\right) }\left( s_{0}^{\pm },s_{1}^{\pm
},...,s_{\Delta k_{\max }-1}^{\pm };0\right) =\left\langle s_{0}^{+}\right|
\rho _{s}\left( 0\right) \left| s_{0}^{-}\right\rangle ,
\end{equation*}%
where%
\begin{eqnarray*}
K\left( s_{k}^{\pm },s_{k+1}^{\pm }\right)  &=&\left\langle
s_{k+1}^{+}\right| \exp (-iH_{0}\Delta t/\hbar )\left|
s_{k}^{+}\right\rangle  \\
&&\times \left\langle s_{k}^{-}\right| \exp (iH_{0}\Delta t/\hbar )\left|
s_{k+1}^{-}\right\rangle .
\end{eqnarray*}%
In the ITM scheme a short-time approximation instead of the Markov
approximation is used. The approximation makes an error of the short-time
propagator in order $\left( \Delta t\right) ^{3},$ which is small enough as
we set the time step $\Delta t$ very small. It is shown that when the time
step $\Delta t$ is not larger than the characteristic time of the qubit
system, which can be calculated with $1/T_{c},$ the calculation is accurate
enough \cite{Privman}. In particular, the scheme does not discard the memory
of the temporal evolution, which may be appropriate to solve the decoherence
of qubit. In the following section we shall use the ITM scheme to study the
decoherence times of the DQD charge qubit in PCPB and DCPB.

\section{Decoherence of DQD charge qubit}

To measure effects of decoherence one can use the entropy, the first
entropy, and many other measures, such as maximal deviation norm, etc. (see
for example Ref. \cite{Privman}). However, essentially, the decoherence of a
open quantum system is reflected through the decays of the off-diagonal
coherent terms of its reduced density matrix. The decoherence is in general
produced due to the interaction of the quantum system with other systems
with a large number of degrees of freedom, for example the devices of the
measurement or environment. The decoherence time denoted by $\tau _{2}$
measures the time of the initial coherent terms to their $1/e$ times,
namely, $\rho _{i}\left( n,m\right) \overset{\tau _{2}}{\rightarrow }\rho
_{f}\left( n,m\right) =\rho _{i}\left( n,m\right) /e.$ Here, $n\neq m,$ and $%
n,$ $m=0$ or $1$ for qubits. In this paper, we investigate the decoherence
times via directly describing the evolutions of the off-diagonal coherent
terms, instead of using any measure of decoherence. In our following
investigations, we suppose the temperature $T=30$ $mK$ and the cut-off
frequency of the bath modes $\omega _{C}=5$ $\left( ps\right) ^{-1}$. We set
the initial state of the qubit to $\rho \left( 0\right) =\frac{1}{2}\left(
\left| 0\right\rangle +\left| 1\right\rangle \right) \left( \left\langle
0\right| +\left\langle 1\right| \right) ,$ which is a pure state and it has
the maximum coherent terms, and the initial state of the environment is $%
\rho _{bath}\left( 0\right) =\prod\nolimits_{k}e^{-\beta M_{k}}/$Tr$%
_{k}\left( e^{-\beta M_{k}}\right) .$ In the calculations we set $\omega
_{d}=0.02$ $(ps)^{-1},$ $T_{c}=0.1\omega _{l}$ according to Ref. \cite%
{Wu-etal}, and two kinds of cases $\omega _{l}=0.5$ $(ps)^{-1}$ and $\omega
_{l}=0.7$ $(ps)^{-1}$ are calculated.

\emph{Decoherence time obtained from ITM scheme:} In the following, at
first, we use the ITM scheme investigating the decoherence time of the DQD
charge qubit. The evolutions of the coherent elements of the reduced density
matrix of the DQD charge qubit in PCPB and DCPB are plotted in Figs. 3 and
4. Here, we simply choose $\Delta k_{\max }=1$ and $\Delta t=1\times 10^{-11}
$ $s$ for PCPB and $\Delta t=2\times 10^{-11}$ $s$ for DCPB in the ITM
scheme. These choices of the time steps are feasible as we consider that it
should be not smaller than the memory times of the baths, because the latter
is about $\tau _{mem}^{pz}\approx 1\times 10^{-11}$ $s$ for PCPB and $\tau
_{mem}^{df}\approx 2\times 10^{-11}$ $s$ for DCPB (see Figs. 1 and 2). It is
also appropriate as we consider that the time steps should not be larger
than the characteristic time of the qubit, because the characteristic time
of the qubit is about $2\times 10^{-11}s$.%
\begin{eqnarray*}
&& \\
&& \\
&&Fig.3, \\
&& \\
&& \\
&&Fig.4 \\
&& \\
&&
\end{eqnarray*}%
Helped with detailed numerical analyses, we can obtain that the decoherence
times of the DQD charge qubit in PCPB are about $\tau _{2}^{pz}\approx 97$ $%
ps$ [when $\omega _{l}=0.7$ $(ps)^{-1}$] and $\tau _{2}^{pz}\approx 118$ $ps$
[when $\omega _{l}=0.5$ $(ps)^{-1}$]. Similarly, we can obtain that the
decoherence times of this qubit in DCPB are about $\tau _{2}^{df}=1.04$ $ps$
[when $\omega _{l}=0.7$ $(ps)^{-1}$] and $\tau _{2}^{df}=3.5$ $ps$ [when $%
\omega _{l}=0.5$ $(ps)^{-1}$]. It is shown that the DCPB behaves more
destructively than the PCPB does to the coherence of the DQD charge qubit. A
further calculation shows that the decoherence time will increase with the
decreasing of $T_{c}.$

\emph{Decoherence time calculated on Bloch equations: }It is well known that
the decoherence time can be calculated based on Bloch equations. In the
following, we calculate the decoherence time of the DQD charge qubit in PCPB
and DCPB with the Bloch equation method. In this method, the relaxation and
dephasing times can be evaluated from the spin-bosonic model with Bloch
equations \cite{Leggett,Weiss}. For our model, they are \cite{BBS}%
\begin{equation*}
\tau _{1}^{-1}=\tau _{2}^{-1}=\frac{1}{2\hbar }J\left( \omega _{0}\right)
\coth \left( \beta \hbar \omega _{0}/2\right) ,
\end{equation*}%
where $\omega _{0}=2T_{c}$ is the natural frequency of the DQD charge qubit.
By using the parameters of the DQD charge qubit and PCPB bath as above, we
can calculate the decoherence times with this method as $\tau
_{2}^{pz}\approx 122.3$ $ps$ [when $\omega _{l}=0.7$ $\left( ps\right) ^{-1}$%
] and $\tau _{2}^{pz}\approx 192.2$ $ps$ [when $\omega _{l}=0.5$ $\left(
ps\right) ^{-1}$]. Similarly, we can obtain the decoherence times of the DQD
charge qubit in the DCPB with this method as $\tau _{2}^{df}\approx 3.18$ $ps
$ [when $\omega _{l}=0.7$ $\left( ps\right) ^{-1}$] and $\tau
_{2}^{df}\approx 12.6$ $ps$ [when $\omega _{l}=0.5$ $\left( ps\right) ^{-1}$%
]. It is shown that the decoherence times obtained from the ITM scheme are
shorter than those obtained on Bloch equations. We suggest that the
differences are derived from the different choices of approximation schemes.
The Bloch equations are in general derived from the Markov approximation
which discards the memory of baths in the derivation of dynamical evolution.
The decoherence of the qubit obtained on Bloch equations is similar to the
``resonant decoherence'' \cite{Openov} obtained from the Fermi golden rule.
It is not accurately equal to the actual decoherence except that the
``nonresonant decoherence'' very small.

\emph{Decoherence time derived from the quality factor:} We like to compare
our results obtained from the ITM scheme based on QUAPI with Thorwart's
results which are also obtained from QUAPI. Thorwart \emph{et al.} \cite%
{Thorwartetal2} investigated the PCPB case and they obtained the quality
factor instead of the decoherence time. By using a set of parameters of the
DQD charge qubit and the PCPB they obtained the quality factor of the qubit
as $Q_{pz}=336,$ which corresponds to decoherence time $\tau
_{2}^{pz}=Q_{pz}\pi /\omega _{pz}^{\prime }\approx 115.9$ $ps,$ where $%
\omega ^{\prime }=\omega _{0}+\Delta \omega ,$ and $\Delta \omega $ is the
bath-induced shift \cite{BBS} in the natural frequency $\omega _{0}=2T_{c}.$
From Fig.1 of Ref. \cite{Wu-etal} we see $\Delta \omega _{pz}\approx
1.75\omega _{c}$ and $\Delta \omega _{df}\approx 1.65\omega _{c}.$ Their
used parameters $[T=10$ $mk,$ $T_{c}\approx 0.07\left( ps\right) ^{-1}]$
have a little difference from ours. But we have calculated that the
difference does not result in much decoherence time departure. It is meant
that our results is in accordance with Thorwart's result. On the other hand,
from our decoherence time $\tau _{2}^{df}\approx 3.5$ $ps$ of the qubit in
DCPB we can obtain its quality factor $Q_{df}=\tau _{2}^{df}\omega
_{df}^{\prime }/\pi \approx 8.$

\section{Discussions and conclusions}

In this paper we investigated the decoherence times of the DQD charge qubit
in PCPB and DCPB with the ITM scheme based on QUAPI. The decoherence times
are also calculated based on Bloch equations. The results derived from the
two kinds of methods are compared to each other. It is shown that the latter
are longer than the former. We think this results from the different choices
of approximation schemes because the Markov approximation used in the latter
method discards the memory of the baths. On the other hand, Hayashi \emph{et
al. }\cite{experiment1,experiment2} have detected that the decoherence time
of the DQD charge qubit is about $1$ $ns$ as $T_{c}\sim 0.07$ $\left(
ps\right) ^{-1}.$ The exact ITM theoretical decoherence times are two orders
of magnitude and five orders of magnitude smaller than the experimental
value even when we consider the DQD charge qubit in independent PCPB and
DCPB. These can finally and without accident lead to the conclusion that the
phonon decoherence is a subordinate mechanism in the DQD charge qubit. In
general, besides the phonon couplings' decoherence, the qubit can also
result in decoherence from electromagnetic fluctuations (with Ohmic noise
spectrum), cotunneling effect, background charge fluctuations (with $1/f$
noise spectrum), and so on. To find out the dominating mechanism of the DQD
charge qubit decoherence and to the best of our abilities to suppress the
central decoherence resources are important challenges in the quantum
computation field.

\begin{acknowledgement}
The project was supported by National Natural Science Foundation of China
(Grant No. 10347133) and Ningbo Youth Foundation (Grant No. 2004A620003).
\end{acknowledgement}

\section{Figure captions}

Fig.1: Real (line) and imaginary (short lines) parts of the response
function of the piezoelectric coupling phonon bath (PCPB). Here, we set the
temperature $T=30$ $mK$, and $\omega _{d}=0.02$ $(ps)^{-1},$ $\omega
_{l}=0.5 $ $(ps)^{-1},$ $g^{pz}=0.035$ $(ps)^{-2}.$

Fig.2: Real (line) and imaginary (short lines) parts of the response
function of the deformation coupling phonon bath (DCPB). Here, we set $%
g^{df}=0.029$ $(ps)^{-2}$, and other parameters are same as those in Fig.1$.$

Fig.3: The evolutions of the off-diagonal elements of the reduced density
matrix for the DQD charge qubit in PCPB when $\omega _{l}=0.5$ $(ps)^{-1}$
(line) and $\omega _{l}=0.7$ $(ps)^{-1}$ (short lines). Here, the cutoff
frequency is $\omega _{c}=5$ $(ps)^{-1}$, other parameters are same as those
in Fig.1. The initial state of the qubit and environment are described in
the text.

Fig.4: The evolutions of the off-diagonal elements of the reduced density
matrix for the DQD charge qubit in DCPB when $\omega _{l}=0.5$ $(ps)^{-1}$
(line) and $\omega _{l}=0.7$ $(ps)^{-1}$ (short lines). Here, the parameters
are same as those in Figs.2 and 3.

\end{document}